# Identification of Bacteria by Patterns Generated from Odor Spectra

Hung-Chih Chang[(+),] Laszlo B. Kish[(+)], Maria D. King[(x)], and Chiman Kwan[(*)]

[(+)]*Department of Electrical and Computer Engineering, Texas A&M University, College Station, TX 77843-3128, USA*

[(x)]*Department of Mechanical Engineering, Texas A&M University, College Station, TX 77843- 3123, USA*

[(*)]*Signal Processing, Inc., 13619 Valley Oak Circle, Rockville, MD 20850, USA*

**Abstract.** We use the power density spectra obtained by fluctuation-enhanced sensing of bacterial odors (*Escherichia coli* and Anthrax-surrogate *Bacillus subtilis*) to generate new, highly distinguishable, types of patterns based on the average slope of the spectra in different frequency ranges. Such plots can be considered as "fingerprints" of bacterial odors. Three different ways of pattern generation are tested, including a simple binary version. The obtained patterns are simple enough to identify the situation by the naked eye without a pattern recognizer.

## 1. Introduction

Bacterium detection and identification has an important role in medical and defense applications. Analyzing their odor [1][2] has good prospects because of high speed, low cost, wide availability, good sensitivity and selectivity, and solid-state electronic noses [3-7] can be applied.

Recently, we have carried out an extensive experimental study [8] with commercial Taguchi sensors to test the shape of the power density spectrum of their electronic noise as an indicator to recognize bacteria. Power density spectrum is one of the easiest and natural tools of Fluctuation-Enhanced Sensing (FES) of chemicals [9-20]. Contrary to simple generation it contains significant sensing information and it has been shown to enhance sensitivity by a factor of 300, or more [14,16]. It is also relatively straightforward to construct a theory to explain its behavior [19].

In the present paper, we develop and test new methods to generate highly distinguishable patterns from the measured spectra. These patterns can be used even by the naked eye or simple algorithms to efficiently identify the bacteria.

In the tests, we use the spectra in paper [8]. For the actual spectra and the details of sample preparation and measurements, see [8] (preprint: http://arxiv.org/abs/0901.3100 ).

**2. Methods for pattern generation from power-density spectra**

In [8] we used the following normalization of the power density spectrum of resistance fluctuations:

$$\gamma(f) = f \frac{S_r(f)}{R_s^2} \quad , \tag{1}$$

$f$ is the frequency, $S_r(f)$ is the power density spectrum of measured resistance fluctuations, and $R_s$ is the measured sensor resistance, and the unit of $\gamma(f)$ is 1. The $\gamma(f)$ for sensors SP32, TGS 2611 and SP11, with the tested bacterium samples, for both the heated and sampling-and-hold sensor measurements, are given in [8], and their amplitudes and slopes have good reproducibility.

Our first step of generating a highly distinguishable pattern is to quantify the average slopes of $\gamma(f)$ in distinct frequency ranges. We measure the average slopes of $\gamma(f)$ in six frequency bands with logarithmically equidistant widths: 100~333Hz, 0.333~1kHz, 1~3.3kHz, 3.3~10kHz, 10~33kHz and 33~100kHz.

Let us make the following notations: $\alpha$ is the average slope in each frequency sub-bands (local slope) and $\beta$ is the average slope over the entire measurement band (100Hz~100KHz), see Figure 1.

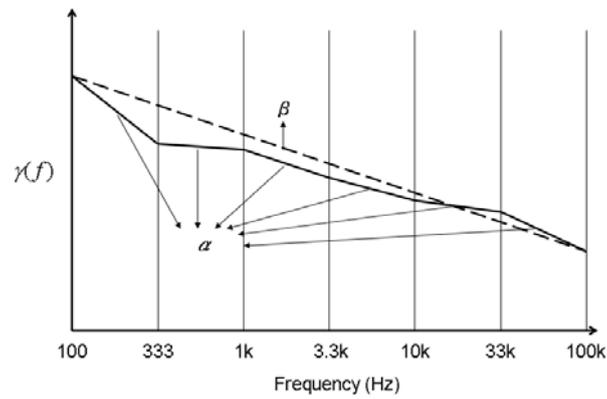

**Figure 1.** Definition of $\alpha$ and $\beta$.

Further the local deviation $\Delta$ is defined for each sub-band as the difference between $\alpha$ and $\beta$ in the following equation

$$\alpha - \beta = \Delta \qquad (2)$$

Finally, the highly distinguishable patterns will be given by the following quantities,

the sign $\sigma$ and the normalized local deviation $\Delta_n$, respectively, see the following definitions:

$$\sigma = \frac{\Delta}{|\Delta|} \tag{3}$$

$$\Delta_n = \log\left|\frac{\Delta}{\beta}\right| \tag{4}$$

Note, the units of both the $\sigma$ and the $\Delta_n$ quantities are one. The quantity $\sigma$ is a binary pattern that indicates if the $\alpha$ is larger or smaller than the $\beta$ for each sub-bands of the spectrum. The advantage of the quantity $\sigma$ is that it provides a very quick way of quantified decision if two spectral patterns are identical or not. Using a simple Boolean logic rule to evaluate these patterns with 6 bit values, can act as a pattern recognizer and distinguish each situation where the pattern is different from the others. On the other hand, $\Delta_n$ is a continuum variable and offers more information when it is needed; however, the quantitative pattern recognition requires more advanced tools.

When using the $\sigma$ histograms to represent our measurement results [8], the binary patterns can be categorized into at least three classes of the four available possibilities, see Figures 2-7, the *empty*, TSA, TSA+ *E. coli* or TSA+ Anthrax, where *empty* means no sample, TSA represents the culture medium: tryptic soy agar, *E. coli* means the harmless laboratory strain MG1655 as a surrogate for the pathogenic vegetative bacteria *Escherichia coli* and Anthrax stands for the anthrax-surrogate bacterium, the sporeforming *Bacillus subtilis*. Moreover, only the sensor SP 32 in the *heated* working mode [8] has distinct sign patterns for the TSA+ *E. coli* and TSA+ Anthrax situations. Therefore, we can conclude that using the simple, binary pattern recognition is able to distinguish the *bacteria* from the *no-bacteria* situations with each sensor in both the *heated* and the *sampling-and-hold* working modes, respectively. However, by using the binary patterns, only the sensor SP 32 working in the *heated* mode is able to distinguish the two bacteria from each other. The conclusions of using binary pattern $\sigma$ are summarized in Table 1.

Figures 8-13 show the plots of the $\Delta_n$ continuum patterns for all different cases. Almost all the different measurement situations (except the *sampling-and-hold* method with SP32) can be categorized into the four different available classes; therefore, the $\Delta_n$ histogram can clearly distinguish all the four types of samples even by the naked eye. However quantified unmanned pattern recognition needs more than a simple logic function to assess the situation. The conclusions of using $\Delta_n$ are summarized in Table 2.

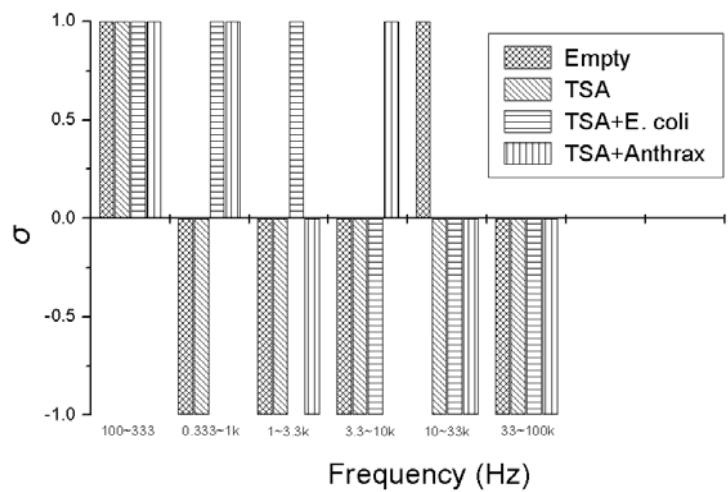

**Figure 2.** The binary pattern $\sigma$ of the heated sensor SP32.

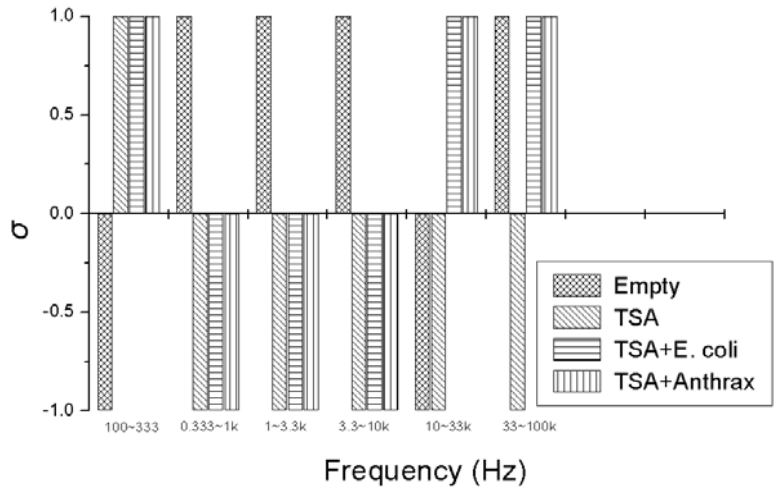

**Figure 3.** The binary pattern $\sigma$ of the sampling-and-hold sensor SP32.

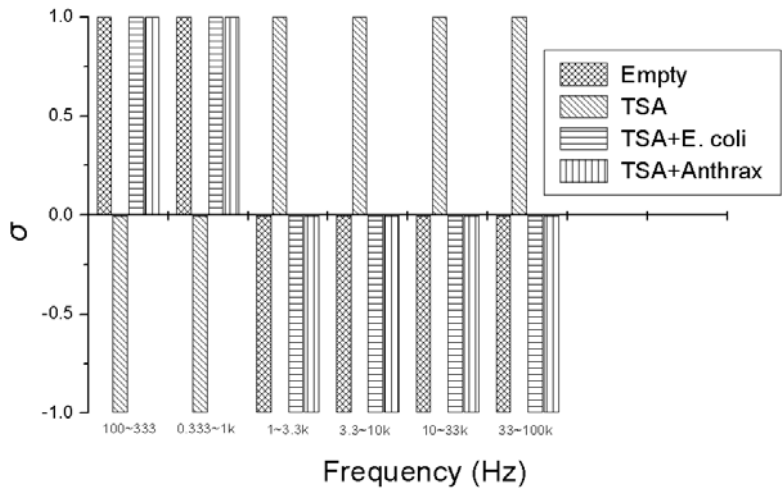

**Figure 4.** The binary pattern $\sigma$ of the heated sensor TGS2611.

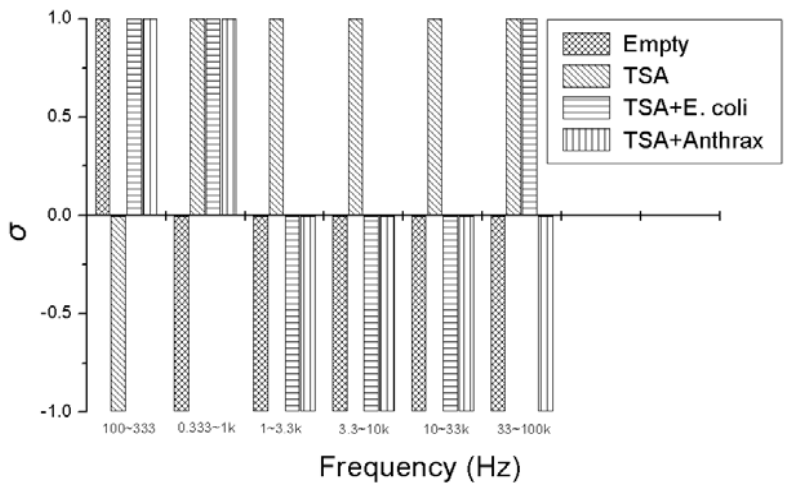

**Figure 5.** The binary pattern $\sigma$ of the sampling-and-hold sensor TGS2611.

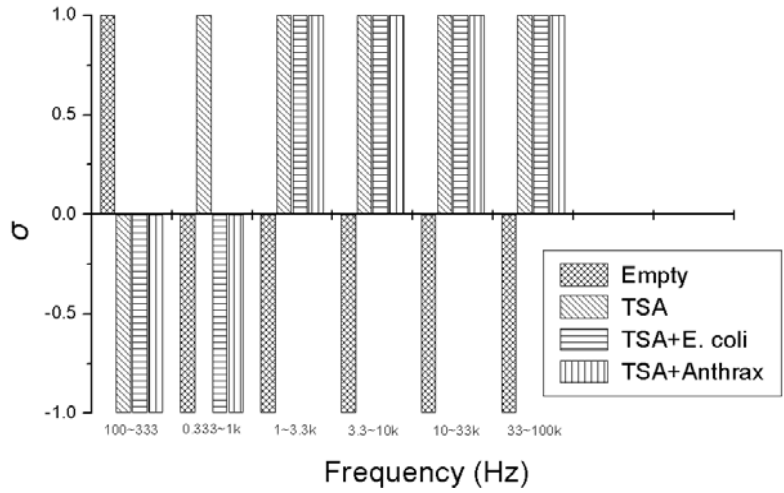

**Figure 6.** The binary pattern $\sigma$ of the heated sensor SP11.

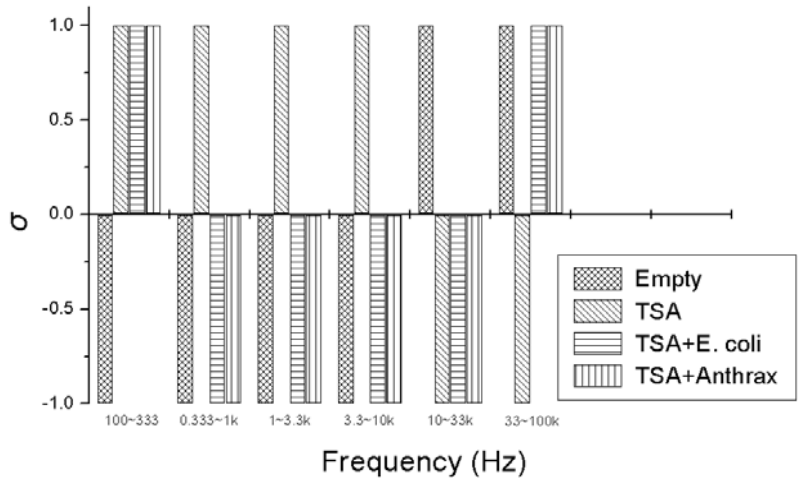

**Figure 7.** The binary pattern $\sigma$ of the sampling-and-hold sensor SP11.

| Sensor | FES Mode | w/o Bacteria | empty/TSA | Bacteria Type |
|--------|----------|--------------|-----------|---------------|
| SP 32 | Heated | + | + | + |
| SP 32 | Sampling-and-hold | + | + | x |
| TGS 2611 | Heated | + | + | x |
| TGS 2611 | Sampling-and-hold | + | + | x |
| SP 11 | Heated | + | + | x |
| SP 11 | Sampling-and-hold | + | + | x |

**Table 1.** Summary table of distinguished samples by using the binary pattern $\sigma$. Notations: + well detected/identified; x unrecognizable.

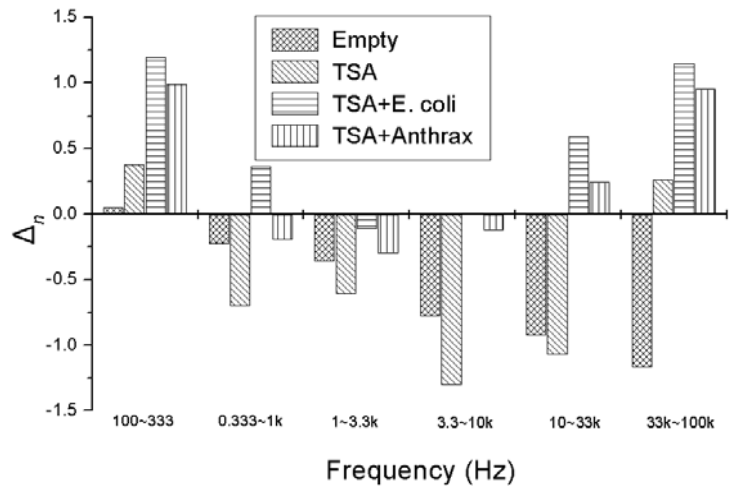

**Figure 8.** The continuum pattern $\Delta_n$ of the heated sensor SP32.

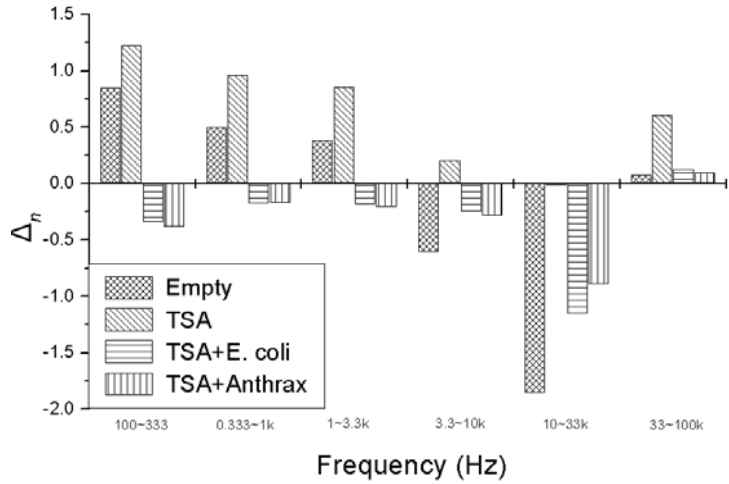

**Figure 9.** The continuum pattern $\Delta_n$ of the sampling-and-hold sensor SP32.

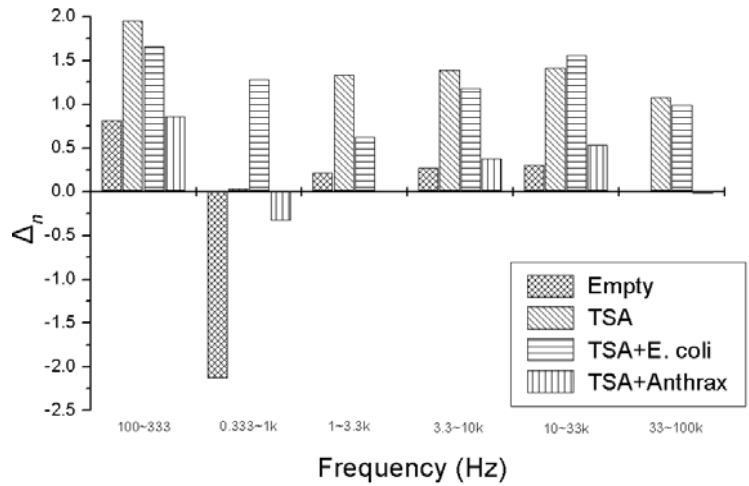

**Figure 10.** The continuum pattern $\Delta_n$ of the heated sensor TGS2611.

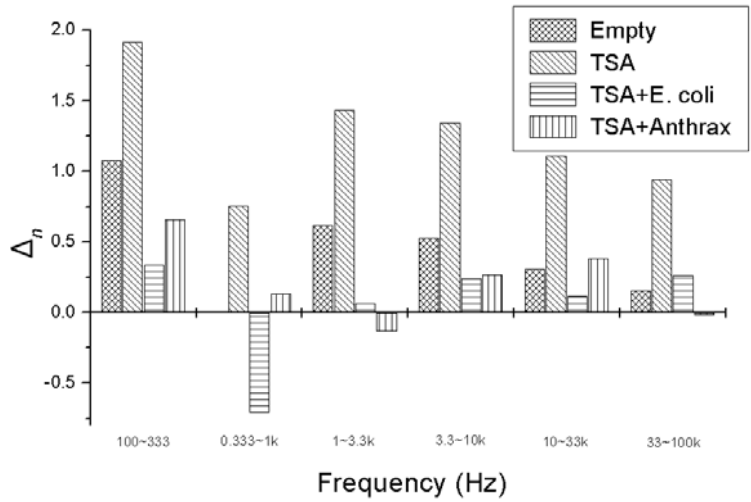

**Figure 11.** The continuum pattern $\Delta_n$ of the sampling-and-hold sensor TGS2611.

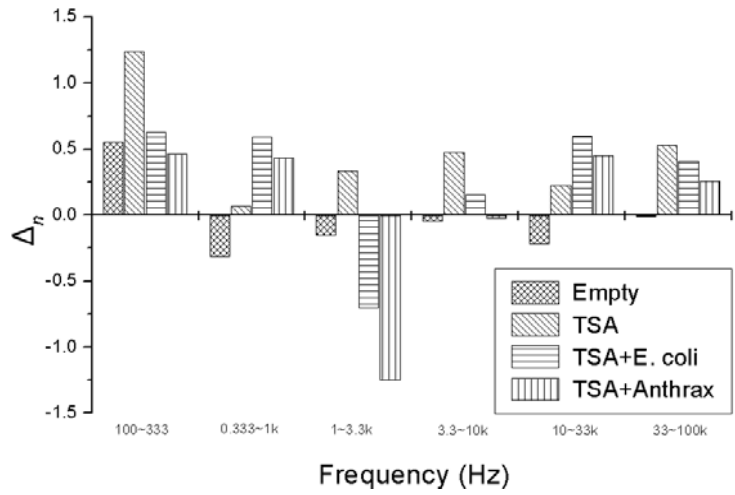

**Figure 12.** The continuum pattern $\Delta_n$ of the heated sensor SP11.

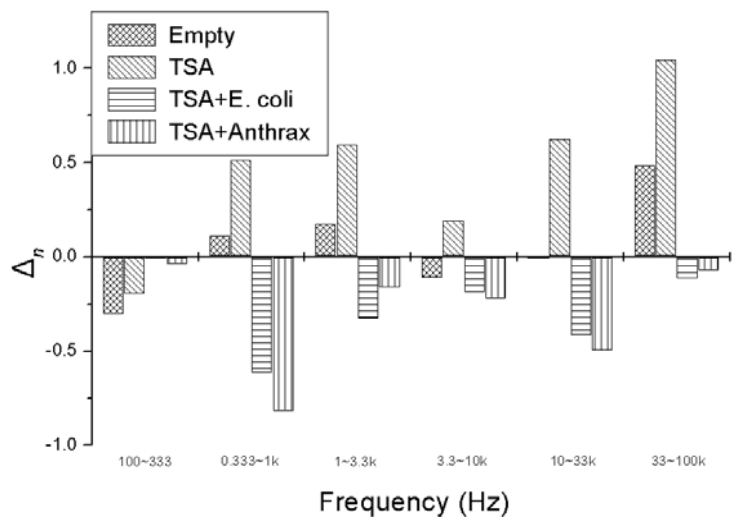

**Figure 13.** The continuum pattern $\Delta_n$ of the sampling-and-hold sensor SP11.

| Sensor | FES Mode | w/o Bacteria | empty/TSA | Bacteria Type |
|---|---|---|---|---|
| SP 32 | Heated | + | + | + |
| SP 32 | Sampling-and-hold | + | + | ? |
| TGS 2611 | Heated | + | + | + |
| TGS 2611 | Sampling-and-hold | + | + | + |
| SP 11 | Heated | + | + | + |
| SP 11 | Sampling-and-hold | + | + | + |

**Table 2.** Summary table of distinguished samples by using the continuum pattern $\Delta_n$. Notations: + well detected/identified; ? at the limit, it may work fine with advanced pattern recognition.

.

## 4. Summary

In this work we have reported an exploratory study to generate and test highly distinguishable types of patterns from power density spectra obtained by fluctuation-enhanced sensing of bacterial odors with single Taguchi sensors. The conclusion of the study is summarized in Tables 1-2.

Further work is required to explore how much the utilization of advanced pattern analysis and spectral decomposition may enhance the performance of these sensors, with these and other types of bacteria, in the fluctuation-enhanced mode.

**Acknowledgment**

This work was supported in part by the Army Research Office under contract W911NF-08-C-0031.

## Biographies

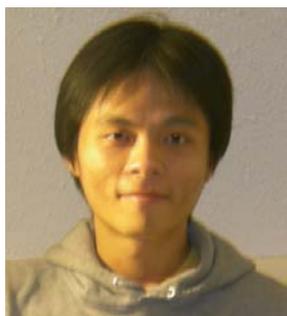**Hung-Chih Chang** was born in Taipei, Taiwan. He received the B.S. degree in electrical engineering from National Central University, Taoyuan, Taiwan in 2000, and the M.S. degree in electrical engineering from National Chiao Tung University, Hsinchu, Taiwan, in 2002. During 2005 and 2006, he worked in TSMC (Taiwan Semiconductor Manufacture Company) in Hsinchu, Taiwan, where he was involved in 90/80/65nm MOS device processes and characterizations. He is currently working towards the Ph.D. degree at the Electrical and Computer Engineering Department of Texas A&M University, College Station. His main research interests include noise and fluctuation, MOSFET modeling, VICOF, chemical and biological sensors.

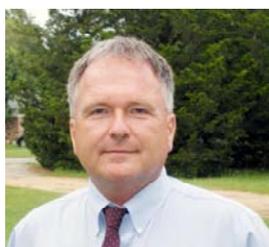**Laszlo B. Kish** (until 1999 Kiss) is full professor at the Electrical and Computer Engineering Department at Texas A&M University, College Station, USA. PhD (physics, JATE Univ., Hungary, 1984), Docent (solid state physics, Uppsala Univ., Sweden 1994), DSc (physics, Hungarian Acad. of Sci., 2001). His main interests are open questions, especially those related to the laws, limits and applications of stochastic fluctuations, noise. He had been and has been working in many related fields including 1/f noise, stochastic resonance, high-Tc superconductors, chemical and biological sensing, and more recently, the error-speed-energy issues of informatics. He is co-inventor of fluctuation-enhanced sensing, and inventor of the unconditionally secure communication via wire with Johnson-like noise, the zero-power communication, the noise-driven computing scenario and the first noise-based logic scheme. He was the founder Editor-in-Chief of Fluctuation and Noise Letters (2001-2008), and founder of symposium series Fluctuations and Noise (SPIE, 2003) and the conference series Unsolved Problems of Noise (1996).

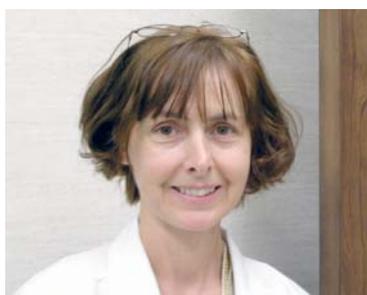**Maria D. King** received her PhD in Microbiology and Chemistry in Germany. She has worked on several microbiology and phage biology related projects that recently led to the development of the prompt bacterial detection technology, the SEPTIC (Sensing of Phage Triggered Ion Cascade). Dr. King is currently conducting biodefense related research studying the effect of bioaerosol collection on the viability and DNA integrity of the aerosolized microorganisms at the Department of Mechanical Engineering, Texas A&M University.

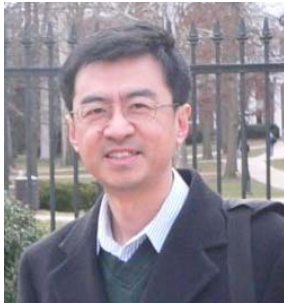**Chiman Kwan** (S'85-M'93-SM'98) received his B.S. degree in electronics with honors from the Chinese University of Hong Kong in 1988 and M.S. and Ph.D. degrees in electrical engineering from the University of Texas at Arlington in 1989 and 1993, respectively.

From April 1991 to February 1994, he worked in the Beam Instrumentation Department of the SSC (Superconducting Super Collider Laboratory) in Dallas, Texas, where he was heavily involved in the modeling, simulation and design of modern digital controllers and signal processing algorithms for the beam control and synchronization system. He received an invention award for his work at SSC. Between March 1994 and June 1995, he joined the Automation and Robotics Research Institute in Fort Worth, where he applied intelligent control methods such as neural networks and fuzzy logic to the control of power systems, robots, and motors. Between July 1995 and March 2006, he was with Intelligent Automation, Inc. in Rockville, Maryland. He served as Principal Investigator/Program Manager for more than sixty five different projects, with total funding exceeding 20 million dollars. Currently, he is the Chief Technology Officer of Signal Processing, Inc., leading research and development efforts in chemical agent detection, biometrics, speech processing, and fault diagnostics and prognostics.

Dr. Kwan's primary research areas include fault detection and isolation, robust and adaptive control methods, signal and image processing, communications, neural networks, and pattern recognition applications. He has published more than 50 papers in archival journals and has had 120 additional papers published in major conference proceedings. He is listed in the New Millennium edition of Who's Who in Science and Engineering and is a member of Tau Beta Pi.